\documentclass[aps,prd,twocolumn,superscriptaddress,notitlepage,nofootinbib,preprintnumbers]{revtex4-1}

\usepackage{epsfig}
\usepackage{multirow}
\usepackage{slashed}
\usepackage{amsmath}
\usepackage{physics}
\usepackage{qcircuit}
\usepackage{braket}
\usepackage{graphicx}
\usepackage{subfigure}
\usepackage{ulem}
\usepackage{color}
\usepackage{epstopdf}
\usepackage[inline]{enumitem}

\def\bea#1\eea{\begin{align}#1\end{align}}
\newcommand{\nn}{\nonumber\\}
\newcommand{\bef}{\begin{figure}[!htp]}
\newcommand{\eef}{\end{figure}}

\usepackage[colorlinks,
linkcolor = blue,
anchorcolor=blue,
citecolor=blue]{hyperref}

\begin{document}
\preprint{RIKEN-iTHEMS-Report-26}
\title{Quantum Simulation of Generalized Parton Distributions in the Schwinger Model}

\date{\today}

\author{Tianyin Li}
\email{tianyin.li@riken.jp}
\affiliation{RIKEN Center for Interdisciplinary Theoretical and Mathematical Sciences (iTHEMS), RIKEN, Wako 351-0198, Japan}

\author{Hongxi Xing}
\email{hxing@m.scnu.edu.cn}
\affiliation{State Key Laboratory of Nuclear Physics and Technology, Institute of Quantum Matter, South China Normal University, Guangzhou 510006, China}
\affiliation{Guangdong Basic Research Center of Excellence for Structure and Fundamental Interactions of Matter, Guangdong Provincial Key Laboratory of Nuclear Science, Guangzhou 510006, China}
\affiliation{Southern Center for Nuclear-Science Theory (SCNT),
Institute of Modern Physics, Chinese Academy of Sciences, Huizhou 516000, China}

\collaboration{QuNu Collaboration}

\date{\today}         

\begin{abstract}
We present a quantum algorithm for simulating Generalized Parton Distributions (GPDs) in the Schwinger model. Unlike the staggered fermions widely utilized in current quantum simulations, we employ Wilson fermions for lattice discretization. This choice is critical for the quantum computation of GPDs due to their strict preservation of charge conjugation symmetry. We construct a comprehensive algorithmic framework that includes the preparation of hadronic states with non-zero momentum and the measurement of light-cone correlation functions incorporating Wilson lines. We provide a complexity analysis, demonstrating that the resources required for our algorithm scale polynomially with both the number of qubits and the desired precision $\varepsilon$. Finally, we benchmark our approach using exact diagonalization, extracting mass spectra and GPDs (also parton distribution functions) that are consistent with theoretical expectations and fundamental physical constraints.

\end{abstract}

\maketitle

\section{Introduction}
A central objective of modern nuclear physics is to elucidate the internal structure of hadrons \cite{Accardi:2012qut,AbdulKhalek:2021gbh,Anderle:2021wcy,CAO:2024fdz}. In general, this hadron structure is encoded by two complementary theoretical pillars: transverse momentum-dependent parton distribution functions (TMD-PDFs) \cite{Boussarie:2023izj} and generalized parton distributions (GPDs) \cite{Diehl:2003ny,Belitsky:2005qn,Ji:2016djn}. While TMD-PDFs provide three-dimensional distributions of partons within the hadron in momentum space, GPDs capture the correlated transverse impact parameter and longitudinal momentum distributions. Together, they provide profound insights into the fundamental hadronic properties, including spin and mechanical structures. In particular, GPDs naturally reduce to standard one-dimensional PDFs in the forward limit, and their first Mellin moments correspond to elastic charge form factors. 

Despite their fundamental importance, calculating GPDs from first principles in quantum chromodynamics (QCD) remains a formidable challenge due to the well-known sign problem. Traditional lattice QCD (LQCD) relies on Euclidean time evolution, which precludes the direct calculation of the real-time, light-cone correlation functions required to define both PDFs and GPDs. Significant theoretical and computational progress has been achieved to bypass this difficulty through large momentum effective theory \cite{Ji:2013dva, Ji:2020ect}, which allows the extraction of light-cone distributions by computing equal-time spatial correlations, known as quasi-PDFs, on a Euclidean lattice. Because these spatial quantities do not suffer from the sign problem, they can be systematically related to the physical light-cone distributions through a perturbative matching process. However, to accurately suppress power corrections and ensure the convergence of the expansion, this approach necessitates simulating hadrons at high momenta, which imposes immense computational demands.

Quantum computing offers a promising paradigm shift to complement these classical lattice methodologies \cite{Fang:2024ple,Zhang:2020uqo,Bauer:2022hpo,Bauer:2023qgm,Banuls:2019bmf}. By directly simulating the Hamiltonian dynamics of gauge theories, quantum devices can theoretically access real-time evolution \cite{Jordan:2012xnu} and evaluate non-local light-cone correlations without encountering the sign problem or requiring extreme momentum boosting \cite{Low:2022jxj}. 
Simulating the GPDs of QCD within a Hamiltonian framework remains a formidable challenge, requiring computational resources that exceed the capabilities of both classical computers and current Noisy Intermediate-Scale Quantum (NISQ) devices. 
As a vital precursor to full (3+1)-dimensional QCD simulations, the Schwinger model, (1+1)-dimensional quantum electrodynamics, serves as an essential and computationally accessible proxy. Despite its reduced dimensionality, the Schwinger model shares critical non-perturbative phenomenological features with QCD, including confinement, mass gap, and chiral symmetry breaking \cite{Banuls:2019bmf}.

In recent years, the community has made significant progress in simulating parton physics utilizing both quantum computing and advanced tensor network methods \cite{Bepari:2021kwv,Li:2022lyt,Barata:2022wim,Li:2023kex,Chawdhry:2023jks,Lee:2023urk,Bauer:2023ujy,Du:2023ewh,Grieninger:2024axp,Li:2024nod,Qian:2024gph,Zhang:2024fgv,Kang:2025xpz,Janik:2025bbz,Banuls:2025wiq,Chen:2025zeh,Chawdhry:2025iuz,Artiaco:2025qqq,Barata:2026icn,Zou:2026cfk,Fernando:2026vqf}. While the primary focus of hadron structure study has been devoted to the study of PDFs, such as extracting PDFs via the quantum simulation of the hadronic tensor \cite{Lamm:2019uyc,Zou:2026cfk,Ikeda:2025bjb}, developing quantum algorithmic frameworks to directly evaluate light-cone correlation functions \cite{Li:2021kcs,Chen:2025zeh}, and utilizing tensor network techniques \cite{Grieninger:2024cdl,Schneider:2024yub,Kang:2025xpz,Banuls:2025wiq}, the quantum computation of GPDs is currently limited to quasi-GPDs in the Schwinger model \cite{Grieninger:2025mbm} and in light-front formalism with leading Fock state truncation \cite{Gustin:2022pfu}.

In this work, we present a comprehensive quantum algorithm to simulate the GPDs of the Schwinger model on a spatial lattice of size $\mathcal{N}$. Because the Schwinger model is restricted to (1+1) dimensions, our calculated GPDs naturally lack the transverse impact-parameter dimensions present in full QCD. Nevertheless, these (1+1)-dimensional GPDs remain highly instructive, allowing for a exploration of the partonic longitudinal momentum fraction ($x$) and the skewness ($\xi$) dependence necessary for describing elastic charge form factors. We extend the scope of previous studies by establishing a unified quantum algorithmic framework capable of computing both PDFs and GPDs, rather than treating them as disjoint evaluations.

A distinctive and critical feature of our approach is the adoption of the Wilson fermion formulation \cite{Wilson:1974sk} for lattice discretization, different with the staggered fermions predominantly employed in current quantum simulations \cite{Zache:2018jbt}. We demonstrate that within the quantum computing framework, implementing Wilson fermions does not incur a higher computational complexity compared to staggered fermions. More importantly, this choice rigorously preserves exact charge conjugation (C), parity (P), and time-reversal (T) symmetries on the lattice. As we will show, the strict preservation of exact C-symmetry is vital for accurately studying the structure of charge-neutral mesons, as it effectively eliminates unphysical real-part contaminations in the correlation functions and guarantees that the resulting GPDs and PDFs remain strictly odd functions with respect to the momentum fraction $x$.

The remainder of this paper is organized as follows. In Sec. \ref{sec-swg}, we introduce the Hamiltonian lattice formulations utilized for the evaluation of GPDs. In Sec. \ref{sec-alg}, we detail the proposed quantum algorithms and provide a rigorous analysis of their circuit complexities. In Sec. \ref{sec-res}, we present the numerical results for the GPDs. Finally, we summarize our findings and discuss future outlooks in Sec. \ref{sec-sum}.

\section{Schwinger model in axial gauge and GPDs}
\label{sec-swg}
\subsection{Schwinger model in Axial gauge}
We begin with a brief review of the Schwinger model in the axial gauge. The continuum Hamiltonian of this model is expressed as:
\bea\label{eq:hax}
    H = & \int d\mathbf{x} \, \bar{\psi}(\mathbf{x})(-i\gamma^1 \partial_1 + m)\psi(\mathbf{x}) + \frac{1}{2} \int d\mathbf{x}\, A^0(\mathbf{x}) J^0(\mathbf{x})\,,
\eea
where $J^\mu(\mathbf{x}) = g\bar{\psi}(\mathbf{x})\gamma^\mu \psi(\mathbf{x})$ is the electric current operator. The temporal gauge field $A^0(\mathbf{x})$ is constrained by Gauss's law:
\bea\label{eq:axgau}
    \partial_1^2 A^0(\mathbf{x}) = -J^0(\mathbf{x})\,.
\eea

To regularize the theory, we discretize the fermion field on a spatial lattice composing $\mathcal{N}$ sites, with lattice spacing $a$ and total volume $\mathcal{N}a$. Crucially, to circumvent the fermion doubling problem while strictly preserving C, P, and T symmetries, we employ Wilson fermions \cite{Wilson:1974sk}. This scheme requires augmenting the Hamiltonian $H$ with a Wilson term $\hat{H}_W$, characterized by a coefficient $r$:
\begin{align}
    \hat{H}_W = \sum_{N=0}^{\mathcal{N}-1}
    \left(-\frac{r}{2} \bar{\psi}(N) \hat{\Delta} \psi(N)\right)\,,
\end{align}
where $\hat{\Delta}$ is the second-order difference operator on the lattice.
For an arbitrary function $f(N)$, this operator is defined as:
\bea\label{eq:latsd}
    \hat{\Delta} f(N) = \frac{1}{a^2} \left[f(N+1) + f(N-1) - 2f(N)\right]\,.
\eea
Incorporating the Wilson term, the Hamiltonian of the lattice Schwinger model becomes \begin{align}\label{eq:disham}
    \hat{H} =& \sum_{N=0}^{\mathcal{N}-1}\left[\frac{1}{2} \bar{\psi}(N) \left(-i\gamma^1-rI\right)\psi(N+1)\right.\nonumber\\
    &\left.+\frac{1}{2}\bar{\psi}(N)(i\gamma^1-rI)\psi(N-1)\right.\nonumber\\
    &\left.+(m+r)\bar\psi(N) \psi(N)+\frac{1}{2}\hat{A}^0(N)\hat{J}^0(N)\right]\,.
\end{align}
The Hamiltonian simplifies significantly if the matrix $(i\gamma^1-rI)$ is diagonal. It is therefore convenient to choose the representation $\gamma^0 = \sigma^x$, $\gamma^1 = -i \sigma^z$, and $\gamma^5 = \gamma^0\gamma^1 = -\sigma^y$, alongside $r=1$. We emphasize that the Hamiltonian in Eq.~\eqref{eq:disham} differs from the Kogut-Susskind formalism \cite{Kogut:1974ag}. Our approach first fixes the theory in the axial gauge before discretizing the fermion and gauge fields on the lattice.

The field $\hat{A}^0(N)$ in the Hamiltonian $\hat{H}$ can be expressed in terms of the fermion field by solving the lattice version of Gauss's law (Eq.~(\ref{eq:axgau})):
\bea\label{eq:lataxgau}
    \hat{\Delta} \hat{A}^0(N) = -\hat{J}^0(N)\,,
\eea
where $\hat{J}^0(N)$ is the charge density operator at spatial site $N$, given by:
\begin{align}
    \hat{J}^0(N) = g\left[\psi^\dagger(N)\psi(N)-1\right]\,.
\end{align}
Solving Eq.~(\ref{eq:lataxgau}) requires specifying a boundary condition. Periodic boundary condition (PBC) is particularly advantageous for calculating GPDs, as they naturally accommodate hadronic states with definite momentum.
Under PBC, the solution for $\hat{A}_0$ is
\begin{align}
    \label{eq:sola0pbc}
    \hat{A}_0(N) =&  \sum_{N^\prime=0}^{\mathcal{N}-1}\left\{\left[{V}(N-N')-{V}(0)\right]\hat{J}_0(N^\prime)\right\}\,,
\end{align}
where the potential $V$ is defined as
\begin{align}
    {V}(N-N') = \sum_{n = 1}^{\mathcal{N}-1} \frac{a}{\mathcal{N}} \frac{e^{i p_n a (N-N')}}{4\sin^2 \frac{p_n a}{2}}\,.
\end{align}
with $p_n = \frac{2\pi n}{\mathcal{N} a}$, $n=0, 1,...,\mathcal{N}-1$. In Eq.~(\ref{eq:sola0pbc}), we subtract ${V}(0)$ to ensure the potential naturally vanishes when $N=N'$.
It can be shown that Eq.~(\ref{eq:sola0pbc}) is a solution to Eq.~(\ref{eq:lataxgau}) in the subspace where the total charge vanishes ($\sum_{N=0}^{\mathcal{N}-1} \hat{J}^0(N) = 0$) and it satisfies the periodicity condition $\hat{A}_0(N) = \hat{A}_0(N+\mathcal{N})$.
Consequently, only the states with zero charge are physical in PBC.

Before mapping the Schwinger model to a quantum register, we arrange the two-component field $\psi_\alpha(N)$ on a $2\mathcal{N}$-qubit chain.
For convenience, the first and second components of the fermion field $\psi$ are assigned to the even and odd sublattices, respectively:
\bea\label{eq:stgg}
    \begin{pmatrix}
    \psi_{1}(N) \\
    \psi_{2}(N) 
    \end{pmatrix}	
    = 
    \begin{pmatrix}
    \phi(2N) \\
    \phi(2N+1) 
    \end{pmatrix}
    \,.
\eea 
We now utilize the single-component field $\phi(n)$, where $0\leq n \leq 2\mathcal{N}-1$, to represent the Schwinger model. This field is mapped to qubits via the Jordan-Wigner transformation \cite{backens_shnirman_makhlin_2019}:
\begin{align}
    \phi(n) = \left(\prod_{n'<n}\sigma^z_{n'}\right)\times\frac{1}{2}\left(\sigma^x_n+i\sigma^y_n \right) .
\end{align}
where $\sigma^i_n$ denotes the Pauli operator acting on the $n$-th qubit. After discretization and Jordan-Wigner transformation, the Hamiltonian in Eq.~(\ref{eq:hax}) can be written as
\bea
    \hat{H} = \hat{H}_1 + \hat{H}_2 + \hat{H}_3 \,,
\eea
where (setting $a=1$)
\begin{align}\label{eq:qubitham}
    \hat{H}_1 =& - \frac{1}{2}\sum_{N={0}}^{\mathcal{N}-1}\left(\sigma^x_{2N+1} \sigma^x_{2N+2} + \sigma^y_{2N+1} \sigma^y_{2N+2}\right) \nonumber\\
    &-\frac{i}{2}(\sigma^x_{2\mathcal{N}-1} \Xi^3_{2\mathcal{N}-1}\sigma^y_0-\sigma^y_{2\mathcal{N}-1} \Xi^3_{2\mathcal{N}-1}\sigma^x_0)\, ,\nonumber\\
    \hat{H}_2 =& \frac{m+1}{2} \sum_{N=0}^{\mathcal{N}-1} (\sigma^x_{2N}\sigma^x_{2N+1}+\sigma^y_{2N}\sigma^y_{2N+1}),& 
    \nonumber \\
    \hat{H}_3 =& \frac{g^2}{4}\sum_{N=0}^{\mathcal{N}-1} \sum_{N'<N}  \left[V(N-N')-V(0)\right]\nonumber\\
    &\times\left(\sum_{\alpha,\beta=0}^{1}\sigma^z_{2N+\alpha} \sigma^z_{2N'+\beta}\right)\,.
\end{align}

\subsection{GPDs in the Schwinger model}
In the Schwinger model, the unpolarized GPDs of a fermion $q$ are defined as the Fourier transform of the gauge-invariant light-cone correlation function  \cite{Collins:1989gx}:
\bea\label{eq:defpdf}
    H_{q/h}(x,\xi) &= \int \frac{dz}{4\pi} e^{-ix z \hat{n}\cdot P} e^{-i z\Delta\cdot \hat n/2}\nn
    &\quad \times \braket{h(p')|\bar{\psi}(z\hat{n}) (\hat n\cdot \gamma) \mathcal{W}(z\hat{n},0)\psi(0)|h(p)}\,,\nn
    &\equiv \int \frac{dz}{4\pi} e^{-ix z \hat{n}\cdot P} \Tilde{H}_{q/h} (z,\xi),
\eea
where $p$ denotes the spacetime momentum of the hadron $h$, $P = (p'+p)/2$, $\Delta = p'-p$ , and $\xi = -\frac{\hat n\cdot \Delta}{2\hat n\cdot P}$ represents the longitudinal momentum transfer of hadron $h$. In Eq.~\eqref{eq:defpdf}, $\psi(z\hat{n})$ represents the fermion field on the light-cone, with $\hat{n}^\mu = (1,1)$. The $\gamma^\mu$ are the standard Dirac gamma matrices, and $\mathcal{W}(z\hat{n},0)$ is the Wilson line, defined as a path-ordered exponential:
\bea\label{eq:defwl}
    \mathcal{W}(z\hat{n},0) = \mathcal{P} \exp[-ig \int_{0}^{z} dy \, \hat{n}^\mu A_\mu(y\hat{n})]\,.
\eea
In the limit $\xi = 0$, GPDs formally reduce to PDFs. In this forward limit, the disconnected part of the light-cone correlation function must be subtracted: 
\bea
    \braket{h(p)|...|h(p)}_c = \braket{h(p)|...|h(p)} - \braket{\Omega|...|\Omega}\,,
\eea
where $\ket{\Omega}$ represents the vacuum state.

\begin{figure*}[htbp]
    \centering    \includegraphics[width=0.90\linewidth]{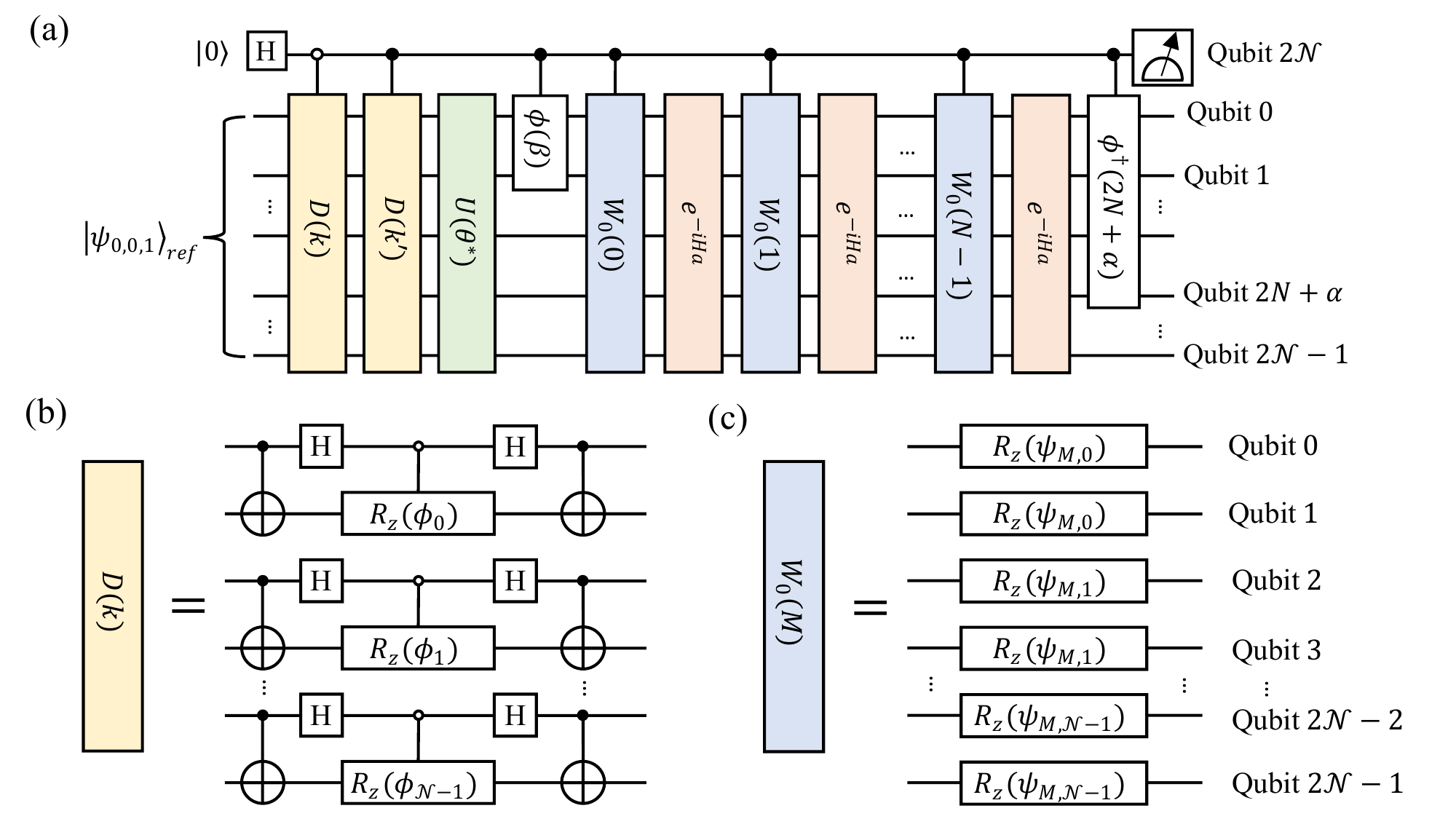}
    \caption{(a) Quantum circuit for simulating GPDs in the Schwinger model.
    The gate $D(k)$ is a circuit that can map $\ket{\psi_{0,0,1}}_{\rm ref}$ to $\ket{\psi_{0,k,0}}_{\rm ref}$.
    The hollow ring over $D(k)$ means $D(k)$ will act on $\ket{\psi_{0,0,1}}_{\rm ref}$ if the Qubits $2\mathcal{N}$ is in $\ket{0}$ state.
    The quantum number resolving VQE $U(\theta)$ is used to prepare the vacuum and hadron states.
    The quantum circuit of $U(\theta)$ will be shown in Fig.~\ref{fig:cir_U}.
    The real and imaginary part of $\tilde{H}^{\alpha \beta}_{q/h}(N)$ can be obtained by measuring the $\sigma^x$ and $\sigma^y$ of the qubit $2\mathcal{N}$. 
    (b) The construction of quantum circuit $D(k)$, with $\phi_{N} =\frac{4\pi N k}{\mathcal{N}}$ if we choose $R_{z}(\theta) = \exp(-i\sigma^z\theta/2)$.
    (c) The quantum circuit for realizing the gauge link $W_0(M)$ in the Schwinger model. By setting $a=1$, the angle $\psi_{M,N} = g^2[V(M-N)-V(0)]$.}
    \label{fig:gpdcir}
\end{figure*}

Following the Wilson fermion discretization, the spatial momentum of the hadronic state evaluates to $\mathbf{p}_k = 2\pi k/\mathcal{N}$, with $k = 0,1,2,...,\mathcal{N}-1$. Thus, a hadron state with special momentum $\mathbf{p}_k$ is denoted as $\ket{h(k)}$ in the Schwinger model. After applying the Jordan-Wigner transformation, the spatial correlation $\tilde{H}_{q/h}(N,\xi)$ in Eq.~(\ref{eq:defpdf}) can be expressed as
\begin{align}\label{eq:latpdf}
    \tilde{H}_{q/h}(N,\xi) &= \sum_{\alpha,\beta=0}^1 (-1)^{\beta} i^{\alpha+\beta} \bra{h(k')} e^{i\hat{H} Na} \phi^\dagger(2N+\alpha) \nonumber\\
    &\quad \times \hat{\mathcal{W}}_{\rm ax}(N,0) \phi(\beta)  \ket{h(k)}\nonumber \\
    &\equiv \sum_{\alpha,\beta=0}^{1} (-1)^\beta i^{\alpha+\beta} \tilde{H}^{\alpha\beta}_{q/h}(N,\xi)\,.
\end{align}
The lattice Wilson line in the axial gauge, $\hat{\mathcal{W}}_{\rm ax}(N,0)$, is constructed as 
\begin{align}\label{eq:wlax}
    \hat{\mathcal{W}}_{\rm ax}(N,0) &= \prod_{N'=0}^{N-1}\left\{e^{-i\hat{H}a} \exp\left[-iga \hat{A}_0(N')\right]\right\}\nonumber\\
    &\equiv \prod_{N'=0}^{N-1} \left[e^{-i \hat{H}a} W_0(N')\right]\,,
\end{align}
where the temporal gauge field evaluates to 
\begin{align}\label{eq:qbA0}
    \hat{A}_0(N) = -\frac{g}{2}\sum_{N'=0}^{\mathcal{N}-1}[V(N-N')-V(0)] \left(\sigma^z_{2N'}+\sigma^z_{2N'+1}\right)\,
\end{align}
under the Jordan-Wigner transformation.

\section{Quantum algorithms for simulating GPDs}
\label{sec-alg}
The quantum simulation of Eq.~\eqref{eq:latpdf} requires the construction of quantum circuits designed to: (1) Prepare the hadronic state $\ket{h}$; (2) Evaluate the gauge-invariant light-cone correlation functions $\tilde{H}^{\alpha \beta}_{q/h}(N,\xi)$.

\subsection{Preparation of hadronic state}

In the lattice Schwinger model with PBC, the good quantum numbers are total electric charge $Q=\sum_{N} \hat{J}^0(N)$, and the spatial momentum $\mathbf{p}_k$. We aim to prepare both the vacuum state $\ket{\Omega}$ and the lightest $q\bar{q}$ bound state $\ket{h(k)}$ across all allowed spatial momentum $\mathbf{p}_k$. All of these target states reside within the uncharged $Q = 0$ physical sector. As demonstrated in Ref. \cite{Li:2021kcs}, these vacuum and hadronic states can be prepared via a quantum-number-resolving variational quantum eigensolver (VQE).

The input states for this algorithm consist  $\mathcal{N}+1$ orthogonal reference states sharing the same quantum number $Q$ and momentum $\mathbf{p}_k$ as target states. We denote these reference states as $\ket{\psi_{Q,k,l}}_{\rm ref}$:
\begin{align}\label{eq:refst}
    \ket{\psi_{0,0,0}}_{\rm ref} &= \otimes_{N=0}^{\mathcal{N}-1}\left[\frac{1}{\sqrt{2}}\left(-\ket{0}_{2N}\otimes \ket{1}_{2N+1}\right.\right.\nonumber\\
    &\left.\left.\quad+\ket{1}_{2N}\otimes \ket{0}_{2N+1}\right)\right]\,, \nonumber\\
    \ket{\psi_{0,k,\delta_{k0}}}_{\rm ref} &= \frac{1}{\sqrt{\mathcal{N}}} \sum_{N=0}^{\mathcal{N}-1} e^{-i \frac{2\pi k N}{\mathcal{N}}}  \ket{\Psi(N)}\,.
\end{align}
where
\begin{align}
    \ket{\Psi(N)} =& \otimes_{M=0}^{\mathcal{N}-1}\left[\frac{1}{\sqrt{2}}\left(-(-1)^{\delta_{NM}}\ket{0}_{2N}\otimes \ket{1}_{2N+1}\right.\right.\nonumber\\
    &\left.\left.\quad+\ket{1}_{2N}\otimes \ket{0}_{2N+1}\right)\right]\,,
\end{align}
and the subscript $2N$ denotes the $2N$-th qubit register. 
The state $\ket{\psi_{0,0,1}}_{\rm ref}$ belongs to a class of Dicke-like states, which can be prepared with an $O(\mathcal{N})$ complexity using the scheme described in Ref. \cite{Li:2022lyt}. 
The reference states $\ket{\psi_{0,k,0}}_{\rm ref}$ are subsequently obtained by applying a momentum-injection circuit $D(k)$ to the state $\ket{\psi_{0,0,1}}_{\rm ref}$:
\begin{align}
    \ket{\psi_{0,k,0}}_{\rm ref} = D(k) \ket{\psi_{0,0,1}}_{\rm ref}\,.
\end{align}

As illustrated in Fig.~\ref{fig:gpdcir}, the operational complexity of $D(k)$ scales as $O(\mathcal{N})$. Therefore, the total computational complexity for preparing the initial reference states remains $O(\mathcal{N})$.

In the quantum-number-resolving VQE framework, a series of trial wave functions is generated by a symmetry-preserving parameterized quantum circuit $U(\theta)$. In the lattice Schwinger model, this variational ansatz $U(\theta)$ is constructed as 
\begin{align}
    U(\theta) = \prod_{j=1}^L\left[ \prod_{i=1}^4 \exp(i\bar{H}_i \theta_{ij})\right]\,,
\end{align}
where $\bar{H}_1 = -\hat{H}_1$, $\bar{H}_2 = 2\hat{H}_3/g^2$, $\bar{H}_3 = -\hat{H}_1$, $\bar{H}_4 = \hat{H}_2/(m+1)$.
To ensure that $U(\theta)$ possesses sufficient expressibility, the number of layers $L$ is chosen to scale linearly with the number of lattice sites, such as $L = \mathcal{N}$.

\begin{figure}
    \centering
    \includegraphics[width=0.98\linewidth]{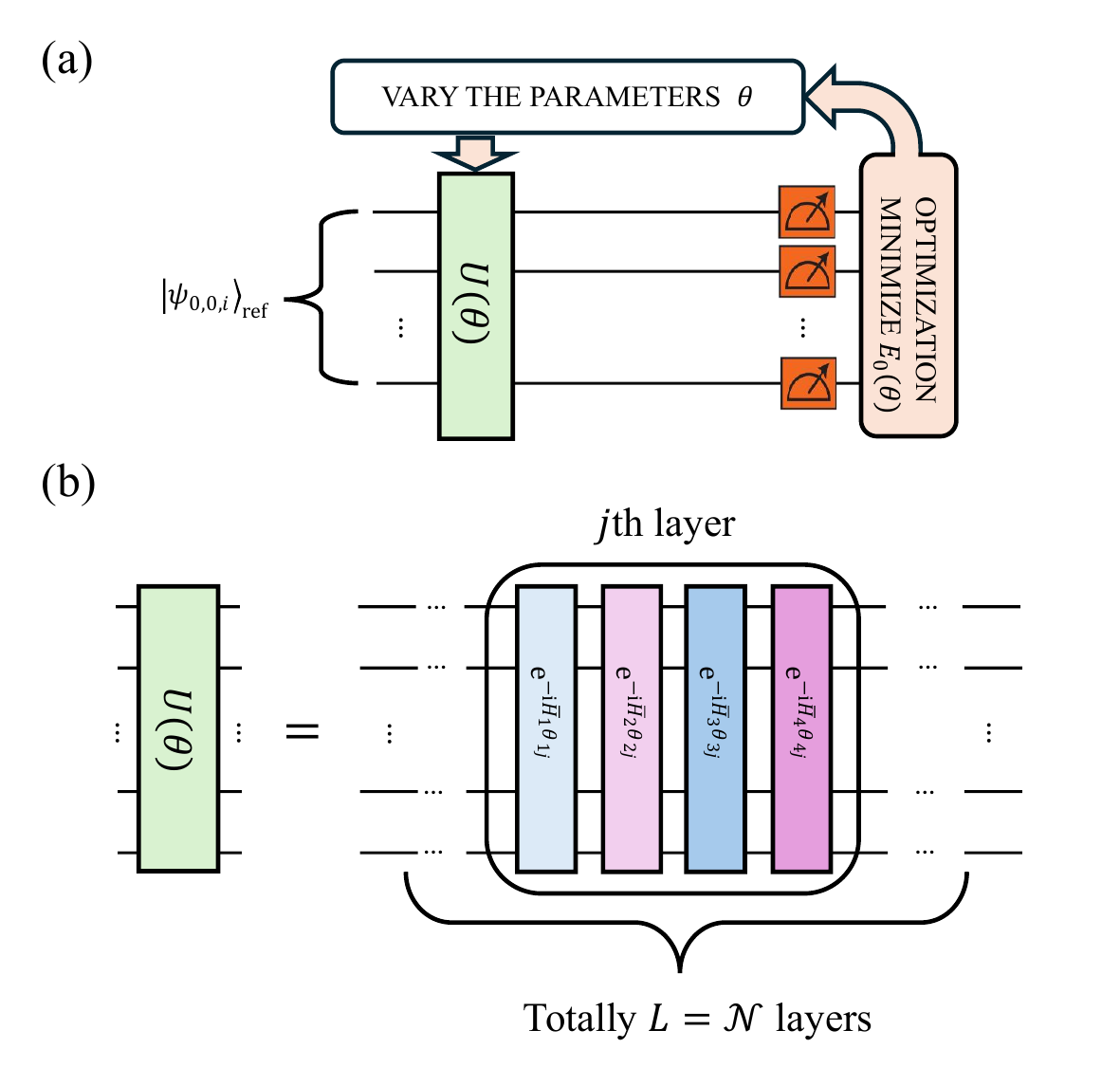}
    \caption{(a).
    The VQE quantum circuit for preparing $\ket{\Omega}$ and $\ket{h}$ states with symmetry-preserving ansatz.
    (b) The construction of the quantum circuit $U(\theta)$.}
    \label{fig:cir_U}
\end{figure}

The quantum circuit of simulating $U(\theta)$ is illustrated in Fig.~\ref{fig:cir_U}.
Within each layer, the dominant computational complexity arises from the simulation of the long-range entangling term $\exp(i \bar{H}_2 \theta_{2j})$. This step carries a complexity of $O(\mathcal{N}^2)$ because the number of non-local interactions in $\bar{H}_2$ scales quadratically as $O(\mathcal{N}^2)$. Therefore, the total complexity for preparing the vacuum and hadronic states scales as $O(\mathcal{N}^2 L)$.

The target states $\ket{\Omega}$ and $\ket{h}$ are optimized simultaneously by minimizing the multi-state loss function
\begin{align}
    E_0(\theta) =& \sum_{i=0}^1 w_{0,0,i} \braket{\psi_{0,0,i}(\theta)|\hat{H}|\psi_{0,0,i}(\theta)}\nonumber\\
    &+\sum_{k=1}^{\mathcal{N}-1} w_{0,k,0}\braket{\psi_{0,k,0}(\theta)|\hat{H}|\psi_{0,k,0}(\theta)}\,,
\end{align}
where the constant weightes satisfy the hierarchy $w_{0,0,1}<w_{0,0,0}$, and we define $\ket{\psi_{Q,k,l}(\theta)} = U(\theta)\ket{\psi_{Q,k,l}}_{\rm ref}$. Denoting the optimal variational parameter set as $\theta^*$, the physical vacuum and hadronic states with arbitrary discrete momenta are prepared via
\begin{align}
    &\ket{\Omega} = U(\theta^*)\ket{\psi_{0,0,0}}_{\rm ref}\,, \nonumber\\
    &\ket{h(k=0)} = U(\theta^*)\ket{\psi_{0,0,1}}_{\rm ref}\,, \nonumber\\
    &\ket{h(k\not=0)} = U(\theta^*)\ket{\psi_{0,k,0}}_{\rm ref}\,.
\end{align}

\subsection{Quantum Simulation of light-cone correlation functions and complexity estimation}
The quantum simulation framework for evaluating non-local correlation functions, originally proposed in Ref. \cite{Pedernales:2014izf}, must be adapted to accommodate two aspects of GPD operators:
(1) The initial and final hadronic states differ due to non-zero momentum transfer; (2) The gauge-invariant operator definitions explicitly incorporate Wilson lines.
The first requirement is fulfilled by implementing a controlled version of the momentum translation operator $D(k)$. 
This controlled gate maps the reference state $\ket{\psi_{0,0,1}}_{\rm ref}$ to a state possessing a well-defined non-vanishing momentum: $\ket{\psi_{0,k,0}}_{\rm ref} = D(k)\ket{\psi_{0,0,1}}_{\rm ref}$. The explicit quantum circuit for $D(k)$ is shown in Fig.~\ref{fig:gpdcir}(b) and requires an $O(\mathcal{N})$ gate count.
To satisfy the second requirement, the local gauge links $W_0(M)$ are controlled by an auxiliary qubit, indexed as the $2\mathcal{N}$-th qubit, to evaluate the gauge link operator $\hat{\mathcal{W}}_{\rm ax}$ defined in Eq.~\eqref{eq:wlax}.
The quantum circuit for $W_0(M)$ is provided in Fig.~\ref{fig:gpdcir}(c) and similarly exhibits a complexity of $O(\mathcal{N})$.

Following the execution of the full quantum circuit shown in Fig.~\ref{fig:gpdcir}(a), the joint density matrix of the system register and the auxiliary qubit is given by:
\begin{align}
    \rho =\frac{1}{2} & (e^{-iHNa}\ket{h(k)}\ket{0}\nonumber\\
    &+\phi^\dagger(2N+\alpha)\hat{\mathcal{W}}_{\rm ax}(N,0)\phi(\beta)\ket{h(k')}\ket{1}) \otimes h.c.
\end{align}
Tracing out the system register fields yields the reduced density matrix of the auxiliary qubit:
\begin{align}
    \rho_A &= \sum_n \bra{n} \rho \ket{n}\nonumber\\
    &= \frac{1}{2} \left[\ket{0} \bra{0}+\ket{1}\bra{1}+\left(\tilde{H}^{\alpha\beta}_{q/h}(N,\xi)\ket{0}\bra{1}+h.c.\right)\right]\,,
\end{align}
where $\ket{n}$ represents the complete basis of eigenstates of the Hamiltonian $\hat{H}$. Consequently, the desired light-cone correlation functions $\tilde{H}^{\alpha\beta}_{q/h}$ can be directly extracted by measuring the off-diagonal expectation values ($\sigma^x$ and $\sigma^y$) of this auxiliary register.

The most complex part in the quantum circuit shown in Fig.~\ref{fig:gpdcir}(a) is the real-time evolution operator $e^{-i\hat Ha}$. This unitary operator can be systematically decomposed into one-qubit and CNOT gates using the second-order Suzuki-Trotter product formula \cite{Childs:2019hts}:
\begin{align}
    e^{-iHa} =& \left[\left(\prod_{j=3}^1e^{-i\hat{H}_j\delta t/2}\right)\left(\prod_{j=1}^3 e^{-i\hat{H}_j \delta t/2}\right)\right]^{N_t}\nonumber\\
    &+\mathcal{O}\left( \frac{a^3}{N_t^2}\right)\,,
\end{align}
where the discrete time step is given by $\delta t = a/N_t$.
The resource requirement for implementing each $e^{-i\hat{H}_j \delta t/2}$ depends on the number of non-vanishing Pauli strings in the corresponding Hamiltonian component $\hat{H}_j$.
As a consequence, simulating the long-range term $e^{-i \hat{H}_3 \delta t}$ requires a complexity of $O(\mathcal{N}^2)$ due to the $\mathcal{N}^2$ scaling of terms in $\hat{H}_3$.

The total number of Trotter steps, $N_t$, is determined by the desired precision $\varepsilon$. Repeating $e^{-iHa}$ for $N$ times leads to the cumulative simulation error by $\mathcal{O}[\varepsilon = Na^3/N_t^2]$. Setting the lattice spacing $a=1$ yields: 
\begin{align}
    N_t = \sqrt{\frac{N}{\varepsilon}} \leq \sqrt{\frac{\mathcal{N}}{\varepsilon}}\,.
\end{align}
From the above discussions, we find that the total complexity for simulating the gauge invariant correlation function $\tilde{H}^{\alpha\beta}_{q/h}(N,\xi)$ scales as $O(\mathcal{N}^2N_t) \leq O(\mathcal{N}^{5/2}/\varepsilon^{1/2})$. Summing the overhead from state preparation and correlation function measurement, the total complexity of the algorithm presented in Fig.~\ref{fig:gpdcir}(a) scales as $O(\mathcal{N}p+\mathcal{N}^{5/2}/\varepsilon^{1/2})$. 
This polynomial scaling with respect to both system size $\mathcal{N}$ and inverse precision $1/\varepsilon$ explicitly demonstrates the advantage of using quantum computers over classical methods to directly simulate GPDs in the Schwinger model.

\section{Numerical results}
\label{sec-res}
It is critical to verify that the axial gauge formalism, specifically the Hamiltonian in Eq.~\eqref{eq:qubitham}, possesses the correct continuum limit. Physical observables, such as the hadron mass, must remain independent of the gauge choice. Thus, the validity of $\hat{H}$ can be tested by calculating its mass spectrum.

\begin{figure}
    \centering
    \includegraphics[width=0.98\linewidth]{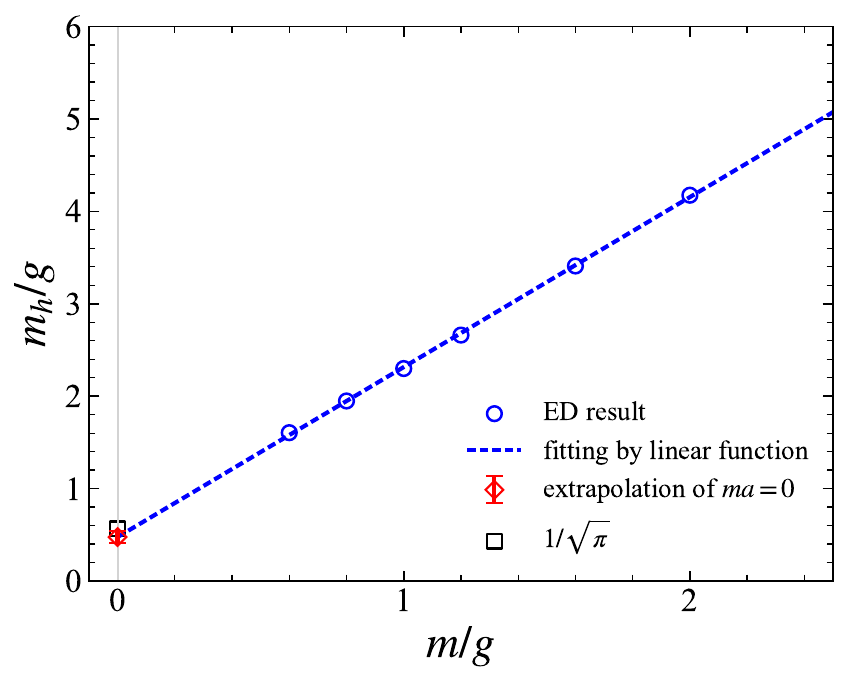}
    \caption{$m/g$ dependence of lightest hadron mass $m_h/g$ of Schwinger model, with fixing $\mathcal{N}=9$ and $m_h a=1.2$.}
    \label{fig:mhmg}
\end{figure}

The mass of the lightest quark-antiquark $q\bar{q}$ hadron in the massless Schwinger model ($m = 0$) was solved exactly in Ref. \cite{Schwinger:1962tp}, yielding $m_h/g = 1/\sqrt{\pi} \approx 0.564$. To reproduce this theoretical limit, we compute $m/g$ dependence of $m_h/g$ at finite $ma$ on the lattice, fixing the system size to $\mathcal{N} = 18$ and lattice spacing $a = 1$.
The exact diagonalization (ED) results for $m_h/g$ are represented by the blue circles in Fig.~\ref{fig:mhmg}, and the massless limit is obtained via linear extrapolation, indicated by the red diamond in Fig.~\ref{fig:mhmg}, which gives $m_h/g = 0.4755 \pm 0.0672$. This extrapolated result from the diagonalizing of $\hat{H}$ is consistent with the analytic calculation, confirming that the Hamiltonian captures the proper continuum limit.

To demonstrate the feasibility of evaluating GPDs within the axial gauge, We calculate the light-cone correlation function $\tilde{H}_{q/h}(N,\xi)$ for the lightest $q\bar{q}$ hadron using the ED method. Because the underlying theory preserves time-reversal symmetry, the relation $\tilde{H}_{q/h}(-N,\xi) = \tilde{H}_{q/h}^*(N,\xi)$ holds. Therefore, the complete behavior of the correlation function is determined strictly from evaluating the $N \geq 0$ domain. 

\begin{figure*}[htbp]
    \centering
    \includegraphics[width=0.98\linewidth]{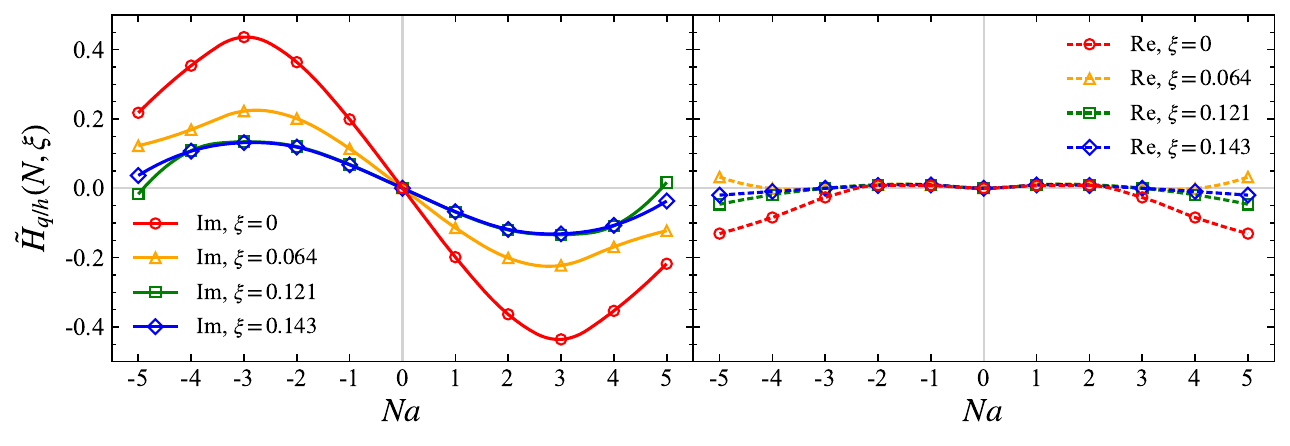}
    \caption{Light-cone correlation function $\tilde{H}_{q/h}(N,\xi)$ of Schwinger model in axial gauge.
    In this figure, we fix $a=1$, $m_h a=1.2$ and the 22 qubits.
    The points data comes from the ED method, while the solid and dashed lines come from the interpolation of data. }
    \label{fig:LFC}
\end{figure*}

\begin{figure}
    \centering
    \includegraphics[width=0.98\linewidth]{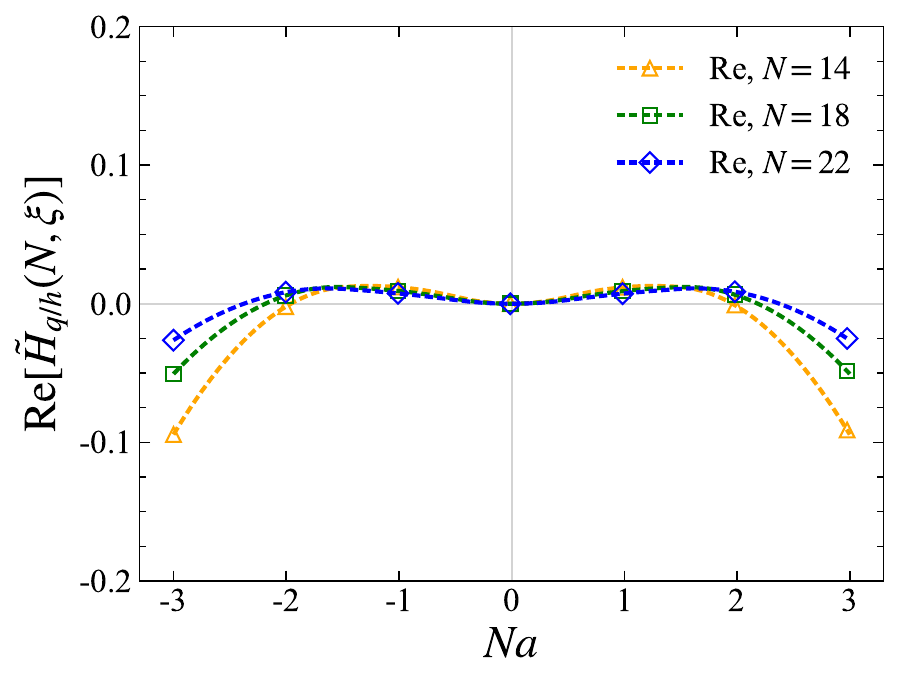}
    \caption{The lattice volume dependance of the real part of correlation function $\tilde{H}_{q/h}(N,\xi)$.
    In this figure, we fix the hadron mass $m_ha = 1.2$, the ratio $m/g = 1$, and the skewness $\xi = 0$.} 
    \label{fig:GPDs_N}
\end{figure}

\begin{figure}[htb]
    \centering
    \includegraphics[width=0.98\linewidth]{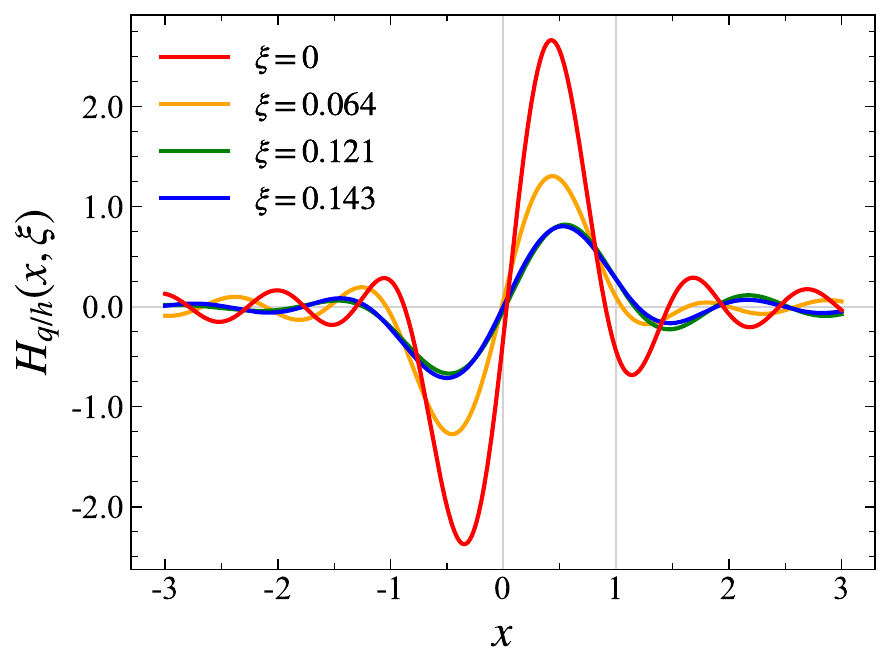}
    \caption{$m_h/g$ dependence of PDFs in Schwinger model.
    This result is obtained by Fourier transforming the light-cone correlation function in Fig.~(\ref{fig:LFC}).}
    \label{fig:GPDs}
\end{figure}

The results for $\tilde{H}_{q/h}$ are depicted in Fig.~\ref{fig:LFC}. The left panel shows the imaginary part of $\tilde{H}_{q/h}$, while right panel displays the real part. The imaginary part ultimately yields the momentum space GPDs, as will be discussed in the context of Fig. \ref{fig:GPDs}. We observe that the real part of $\tilde{H}_{q/h}$ exhibits small but none-zero values. In the exact continuum limit, this real part must vanish since the states $\ket{h(k)}$ are eigenstates of the charge conjugation operator $\mathcal{C}$. Although the Wilson fermion formalism explicitly preserves $C$-symmetry, the non-vanishing real part of $\tilde{H}_{q/h}$ at large spatial $N$ region emerges as a finite-volume artifacts. Under PBC, the spatial geometry is effectively circular. At large spatial separations $z$, the extended Wilson line becomes sensitive to this finite topology. Consequently, as the total lattice volume is increased, the real part of $\tilde{H}_{q/h}$ tends asymptotically toward zero. This expected lattice volume dependence is verified in Fig.~\ref{fig:GPDs_N}, which confirms that the real part of $\tilde{H}_{q/h}$ systematically decreases as the lattice volume increases.

\begin{figure}
    \centering
    \includegraphics[width=0.98\linewidth]{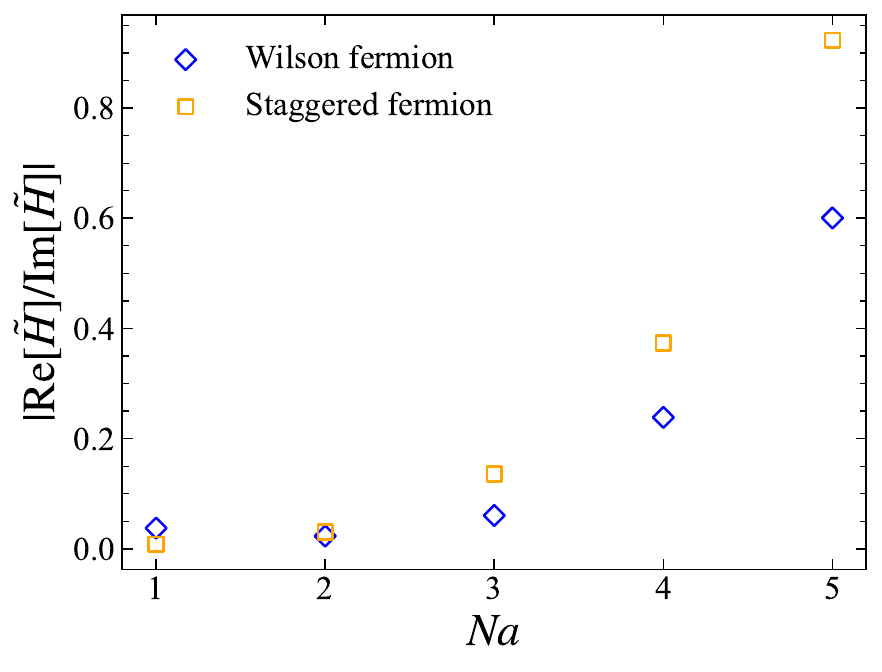}
    \caption{The ration of real and imaginary part of $\tilde{H}_{q/h}(N>0,\xi=0)$. The orange squre data point is calculated by the Staggered fermion formalism while the blue diamon data point is calculated by the Wilson fermion method.}
    \label{fig:ST_WF}
\end{figure}

In contrast, realizing $C$-symmetry within the staggered fermion formalism significantly more complicated. As a result, $\tilde{H}_{q/h}(N, \xi)$ calculated using staggered fermions exhibits a more pronounced non-zero real part, which stems from a combination of finite-volume effects and the explicit violation of exact $C$-symmetry. To quantify the magnitude of this contamination, we evaluate the ratio $|\text{Re}(\tilde{H}_{q/h})/\text{Im}(\tilde{H}_{q/h})|$. 
Fig.~\ref{fig:ST_WF} compares this ratio for Wilson fermions (blue diamonds) and staggered fermions (orange squares) with fixed parameters $N=9$, $m_h a = 1.2$, $m/g = 1$, and $\xi=0$. 
The ratio is significantly suppressed for Wilson fermions compared to their staggered counterparts. This comparative advantage demonstrates that the Wilson fermion formalism yields superior performance when calculating GPDs (also PDFs) in the Schwinger model, an improvement primarily attributable to its rigorous preservation of $C$-symmetry.

The momentum space GPDs ${H}_{q/h}(x,\xi)$ of the Schwinger model are obtained via the Fourier transform of the spatial light-cone correlation functions $\tilde{H}_{q/h}(N,\xi)$ shown in Fig.~\ref{fig:LFC}. To generate continuous and smooth GPD profiles, we first interpolate the discrete lattice data points and subsequently apply the transformation to the resulting interpolation curves. The final evaluated GPDs are shown in Fig.~\ref{fig:GPDs}. 
In the limit of $\xi=0$, the GPD $H_{q/h}(x,\xi=0)$ reduces to the standard PDF. The peak of this PDFs occurs around $x = 0.5$, correctly reflecting that the lightest $q\bar{q}$ bound state predominantly consists of two valence quarks carrying equal momentum fraction in 1+1 dimension. As the momentum transfer $\xi$ increases, the peak of $\tilde{H}_{q/h}(x,\xi)$ shifts toward larger $x$-values while simultaneously becoming broader and more suppressed. This qualitative behavior aligns well with recent tensor network calculations of the GPDs in the Schwinger model \cite{Grieninger:2025mbm}.

Furthermore, the extracted GPDs are odd functions of $x$. This is an expected theoretical property because the elastic charge form factor $F(\xi)$ is given by the integral of GPDs over $x$. Since the $q\bar{q}$ bound state considered here is an eigenstate of the charge conjugation operator, its elastic form factor must strictly vanish: 
\begin{align}
    F(\xi) = \int dx \tilde{H}_{q/h}(x,\xi) = 0\,.
\end{align}
Consequently, the corresponding GPDs must remain anti-symmetric (odd) functions for any arbitrary non-zero momentum transfer $\xi$, which further confirms our result as shown in Fig. \ref{fig:GPDs}.

\section{Summary}
\label{sec-sum}
In this work, we have presented a comprehensive quantum algorithm designed to GPDs of the Schwinger model within the axial gauge using Wilson fermion formalism. The direct evaluation of GPDs requires the preparation of physical hadronic bound states with well-defined finite momentum, and the evaluation of non-local, gauge-invariant light-cone correlation functions. We prepare the hadron states by implementing a symmetry-preserving, quantum-number-resolving VQE, which reliably prepares the required momentum-boosted target states without breaking the underlying symmetries of the Hamiltonian. We systematically estimated the total computational complexity of our proposed quantum algorithm, which scales as $O(\mathcal{N}p+\mathcal{N}^{5/2}/\varepsilon^{1/2})$. This polynomial scaling explicitly demonstrates the theoretical feasibility and advantage of utilizing quantum computing architectures.

To validate our theoretical framework, we employed exact diagonalization (ED) methods to explicitly calculate the bound state mass spectrum and the corresponding GPDs of the Schwinger model. The extracted mass spectrum reproduced the known analytical continuum limits. Furthermore, by evaluating the spatial correlation functions and calculating their subsequent Fourier transforms, we obtained the momentum-space GPDs. Our comparative results underscored the critical importance of preserving exact charge conjugation symmetry. Specifically, we demonstrated that the Wilson fermion formalism significantly suppresses unphysical components compared to the staggered fermion approach.

Looking forward, the methodologies established in this paper provide a framework for exploring more complex quantum field theories. A natural and immediate extension of this work is the generalization of our algorithm to calculate the GPDs of non-Abelian gauge theories in (1+1) dimensions, such as SU(2) or SU(3) Yang-Mills theories coupled to fermions. Investigating these non-Abelian models will serve as a crucial step toward the ultimate goal of simulating full (3+1)-dimensional QCD on future fault-tolerant quantum hardware. 

\begin{acknowledgments}
We thank QuNu members and Jian Zhou for valuable discussions. This work is supported by the National Natural Science Foundation of China with Project Nos. 12525508, 12475139.
\end{acknowledgments}

\normalem

\begin{thebibliography}{10}

\bibitem{Accardi:2012qut}
A.~Accardi {\em et~al.},
\newblock Eur. Phys. J. A {\bf 52}, 268 (2016), arXiv:1212.1701.

\bibitem{AbdulKhalek:2021gbh}
R.~Abdul~Khalek {\em et~al.},
\newblock Nucl. Phys. A {\bf 1026}, 122447 (2022), arXiv:2103.05419.

\bibitem{Anderle:2021wcy}
D.~P. Anderle {\em et~al.},
\newblock Front. Phys. (Beijing) {\bf 16}, 64701 (2021), arXiv:2102.09222.

\bibitem{CAO:2024fdz}
X.~CAO {\em et~al.},
\newblock Nucl. Tech. {\bf 43}, 020001 (2024).

\bibitem{Boussarie:2023izj}
R.~Boussarie {\em et~al.},
\newblock (2023), arXiv:2304.03302.

\bibitem{Diehl:2003ny}
M.~Diehl,
\newblock Phys. Rept. {\bf 388}, 41 (2003), arXiv:hep-ph/0307382.

\bibitem{Belitsky:2005qn}
A.~V. Belitsky and A.~V. Radyushkin,
\newblock Phys. Rept. {\bf 418}, 1 (2005), arXiv:hep-ph/0504030.

\bibitem{Ji:2016djn}
X.~Ji,
\newblock Natl. Sci. Rev. {\bf 4}, 213 (2017), arXiv:1605.01114.

\bibitem{Ji:2013dva}
X.~Ji,
\newblock Phys. Rev. Lett. {\bf 110}, 262002 (2013), arXiv:1305.1539.

\bibitem{Ji:2020ect}
X.~Ji, Y.-S. Liu, Y.~Liu, J.-H. Zhang, and Y.~Zhao,
\newblock Rev. Mod. Phys. {\bf 93}, 035005 (2021), arXiv:2004.03543.

\bibitem{Fang:2024ple}
Y.~Fang {\em et~al.},
\newblock Sci. China Phys. Mech. Astron. {\bf 68}, 260301 (2025),
  arXiv:2411.11294.

\bibitem{Zhang:2020uqo}
D.-B. Zhang, H.~Xing, H.~Yan, E.~Wang, and S.-L. Zhu,
\newblock Chin. Phys. B {\bf 30}, 020306 (2021), arXiv:2011.01431.

\bibitem{Bauer:2022hpo}
C.~W. Bauer {\em et~al.},
\newblock PRX Quantum {\bf 4}, 027001 (2023), arXiv:2204.03381.

\bibitem{Bauer:2023qgm}
C.~W. Bauer, Z.~Davoudi, N.~Klco, and M.~J. Savage,
\newblock Nature Rev. Phys. {\bf 5}, 420 (2023), arXiv:2404.06298.

\bibitem{Banuls:2019bmf}
M.~C. Ba{\~n}uls {\em et~al.},
\newblock Eur. Phys. J. D {\bf 74}, 165 (2020), arXiv:1911.00003.

\bibitem{Jordan:2012xnu}
S.~P. Jordan, K.~S.~M. Lee, and J.~Preskill,
\newblock Science {\bf 336}, 1130 (2012), arXiv:1111.3633.

\bibitem{Low:2022jxj}
G.~H. Low, Y.~Su, Y.~Tong, and M.~C. Tran,
\newblock PRX Quantum {\bf 4}, 020323 (2023), arXiv:2211.09133.

\bibitem{Bepari:2021kwv}
K.~Bepari, S.~Malik, M.~Spannowsky, and S.~Williams,
\newblock Phys. Rev. D {\bf 106}, 056002 (2022), arXiv:2109.13975.

\bibitem{Li:2022lyt}
QuNu, T.~Li {\em et~al.},
\newblock Sci. China Phys. Mech. Astron. {\bf 66}, 281011 (2023),
  arXiv:2207.13258.

\bibitem{Barata:2022wim}
J.~Barata, X.~Du, M.~Li, W.~Qian, and C.~A. Salgado,
\newblock Phys. Rev. D {\bf 106}, 074013 (2022), arXiv:2208.06750.

\bibitem{Li:2023kex}
QuNu, T.~Li, W.~K. Lai, E.~Wang, and H.~Xing,
\newblock Phys. Rev. D {\bf 109}, 036025 (2024), arXiv:2301.04179.

\bibitem{Chawdhry:2023jks}
H.~A. Chawdhry and M.~Pellen,
\newblock SciPost Phys. {\bf 15}, 205 (2023), arXiv:2303.04818.

\bibitem{Lee:2023urk}
K.~Lee, J.~Mulligan, F.~Ringer, and X.~Yao,
\newblock Phys. Rev. D {\bf 108}, 094518 (2023), arXiv:2308.03878.

\bibitem{Bauer:2023ujy}
C.~W. Bauer, S.~Chigusa, and M.~Yamazaki,
\newblock Phys. Rev. A {\bf 109}, 032432 (2024), arXiv:2310.19881.

\bibitem{Du:2023ewh}
X.~Du and W.~Qian,
\newblock Phys. Rev. D {\bf 109}, 076025 (2024), arXiv:2312.16294.

\bibitem{Grieninger:2024axp}
S.~Grieninger and I.~Zahed,
\newblock Phys. Rev. D {\bf 110}, 116009 (2024), arXiv:2406.01891.

\bibitem{Li:2024nod}
T.~Li, H.~Xing, and D.-B. Zhang,
\newblock (2024), arXiv:2406.05683.

\bibitem{Qian:2024gph}
W.~Qian, M.~Li, C.~A. Salgado, and M.~Kreshchuk,
\newblock Phys. Rev. D {\bf 111}, 096001 (2025), arXiv:2411.09762.

\bibitem{Zhang:2024fgv}
G.~Zhang, X.~Guo, E.~Wang, and H.~Xing,
\newblock Phys. Rev. D {\bf 111}, 056031 (2025), arXiv:2411.18869.

\bibitem{Kang:2025xpz}
Z.-B. Kang, N.~Moran, P.~Nguyen, and W.~Qian,
\newblock JHEP {\bf 09}, 176 (2025), arXiv:2501.09738.

\bibitem{Janik:2025bbz}
R.~A. Janik, M.~A. Nowak, M.~M. Rams, and I.~Zahed,
\newblock Phys. Rev. Lett. {\bf 135}, 211903 (2025), arXiv:2502.12901.

\bibitem{Banuls:2025wiq}
M.~C. Ba{\~n}uls, K.~Cichy, C.~J.~D. Lin, and M.~Schneider,
\newblock (2025), arXiv:2504.07508.

\bibitem{Chen:2025zeh}
J.-W. Chen, Y.-T. Chen, and G.~Meher,
\newblock (2025), arXiv:2506.16829.

\bibitem{Chawdhry:2025iuz}
H.~A. Chawdhry, M.~Pellen, and S.~Williams,
\newblock (2025), arXiv:2507.07194.

\bibitem{Artiaco:2025qqq}
C.~Artiaco, J.~Barata, and E.~Rico,
\newblock (2025), arXiv:2510.16101.

\bibitem{Barata:2026icn}
J.~Barata, M.~Li, W.~Qian, C.~A. Salgado, and J.~M. Silva,
\newblock (2026), arXiv:2604.11616.

\bibitem{Zou:2026cfk}
D.~Zou, T.~Li, J.~Liang, E.~Wang, and H.~Xing,
\newblock (2026), arXiv:2606.17003.

\bibitem{Fernando:2026vqf}
I.~P. Fernando and D.~Keller,
\newblock (2026), arXiv:2604.10025.

\bibitem{Lamm:2019uyc}
NuQS, H.~Lamm, S.~Lawrence, and Y.~Yamauchi,
\newblock Phys. Rev. Res. {\bf 2}, 013272 (2020), arXiv:1908.10439.

\bibitem{Ikeda:2025bjb}
K.~Ikeda, Z.-B. Kang, D.~E. Kharzeev, and W.~Qian,
\newblock (2025), arXiv:2512.18062.

\bibitem{Li:2021kcs}
QuNu, T.~Li {\em et~al.},
\newblock Phys. Rev. D {\bf 105}, L111502 (2022), arXiv:2106.03865.

\bibitem{Grieninger:2024cdl}
S.~Grieninger, K.~Ikeda, and I.~Zahed,
\newblock Phys. Rev. D {\bf 110}, 076008 (2024), arXiv:2404.05112.

\bibitem{Schneider:2024yub}
M.~Schneider, M.~C. Ba{\~n}uls, K.~Cichy, and C.~J.~D. Lin,
\newblock PoS {\bf LATTICE2024}, 024 (2025), arXiv:2409.16996.

\bibitem{Grieninger:2025mbm}
S.~Grieninger, J.~Montgomery, F.~Ringer, and I.~Zahed,
\newblock (2025), arXiv:2511.17752.

\bibitem{Gustin:2022pfu}
C.~M. Gustin and G.~Goldstein,
\newblock (2022), arXiv:2211.07826.

\bibitem{Wilson:1974sk}
K.~G. Wilson,
\newblock Phys. Rev. D {\bf 10}, 2445 (1974).

\bibitem{Zache:2018jbt}
T.~V. Zache {\em et~al.},
\newblock Quantum Sci. Technol. {\bf 3}, 034010 (2018), arXiv:1802.06704.

\bibitem{Kogut:1974ag}
J.~B. Kogut and L.~Susskind,
\newblock Phys. Rev. D {\bf 11}, 395 (1975).

\bibitem{backens_shnirman_makhlin_2019}
S.~Backens, A.~Shnirman, and Y.~Makhlin,
\newblock Scientific Reports {\bf 9} (2019).

\bibitem{Collins:1989gx}
J.~C. Collins, D.~E. Soper, and G.~F. Sterman,
\newblock Adv. Ser. Direct. High Energy Phys. {\bf 5}, 1 (1989),
  arXiv:hep-ph/0409313.

\bibitem{Pedernales:2014izf}
J.~{\,}. Pedernales, R.~Di~Candia, I.~{\,}. Egusquiza, J.~Casanova, and
  E.~Solano,
\newblock Phys. Rev. Lett. {\bf 113}, 020505 (2014), arXiv:1401.2430.

\bibitem{Childs:2019hts}
A.~M. Childs, Y.~Su, M.~C. Tran, N.~Wiebe, and S.~Zhu,
\newblock Phys. Rev. X {\bf 11}, 011020 (2021), arXiv:1912.08854.

\bibitem{Schwinger:1962tp}
J.~S. Schwinger,
\newblock Phys. Rev. {\bf 128}, 2425 (1962).

\end{thebibliography}

\end{document}